\documentclass[10pt,conference,compsoc]{IEEEtran}
\usepackage[utf8]{inputenc}
\usepackage{amsmath}
\usepackage{amssymb}
\usepackage{graphicx}
\usepackage{listings}
\usepackage{xcolor}
\usepackage{booktabs}
\usepackage{array}
\usepackage[
    colorlinks=true,
    urlcolor=blue,
    linkcolor=blue
]{hyperref}
\definecolor{codegreen}{rgb}{0,0.6,0}
\definecolor{codegray}{rgb}{0.5,0.5,0.5}
\definecolor{codepurple}{rgb}{0.58,0,0.82}
\definecolor{backcolour}{rgb}{0.95,0.95,0.92}

\lstdefinestyle{mystyle}{
    backgroundcolor=\color{backcolour},   
    commentstyle=\color{codegreen},
    keywordstyle=\color{magenta},
    numberstyle=\tiny\color{codegray},
    stringstyle=\color{codepurple},
    basicstyle=\ttfamily\footnotesize,
    breakatwhitespace=false,         
    breaklines=true,                 
    captionpos=b,                    
    keepspaces=true,                 
    numbers=left,                    
    numbersep=5pt,                  
    showspaces=false,                
    showstringspaces=false,
    showtabs=false,                  
    tabsize=2
}
\title{TITAN-FedAnil+: Trust-Based Adaptive Blockchain Federated Learning for Resource-Constrained Intelligent Enterprises}

\author{
\IEEEauthorblockN{Muhammad Hadi, Muhammad Jahangir, Talha Shafique, and Muhammad Khuram Shahzad}
\IEEEauthorblockA{School of Electrical Engineering and Computer Science (SEECS)\\
National University of Sciences and Technology (NUST)\\
Islamabad, Pakistan\\
Email: \{mhadi, mjahangir, tshafique\}.mscs25seecs@seecs.edu.pk\\
Corresponding author: Muhammad Khuram Shahzad (khuram.toor@gmail.com)}
}

\begin{document}

\maketitle

\begin{abstract}
Federated Learning (FL) has proven to be a central approach to achieving collaborative intelligence without compromising privacy. Data heterogeneity (non-IID distribution) and decentralized security, however, are still formidable challenges particularly in resource-limited environments. We introduce TITAN-FedAnil+, a \textbf{T}rust-based \textbf{A}daptive \textbf{N}etwork for blockchain-based federated learning in intelligent enterprises. We use Affinity Propagation (AP) based adaptive clustered aggregation as a method to filter malicious updates without knowing the number of attackers. Additionally, we propose GPU vectorization and ``Signed State Jumps'' to enable efficient resynchronization, which brings memory overhead reduction to a peak of \textbf{81.00\%} in 50 communication rounds on highly constrained 8GB edge networks as compared to catastrophic performance by the baseline model at Round 3.\\
The code for the project is available at:
\href{https://github.com/error8149/FedAnilPlus-Optimized}
{\textbf{TITAN-FedAnil+}}
\end{abstract}


\section{Introduction}

The Internet of Things (IoT) has seen a dramatic increase in the number of devices and the development of Intelligent Enterprises (IEs) has sparked a surge in the drive to make decisions based on data. Collecting such sensitive enterprise information at a central point in vast amounts in the cloud, however, creates serious privacy concerns, regulatory issues (such as GDPR), and major communication bottlenecks \cite{fedavg}. One solution that is gaining traction is called Federated Learning (FL) where decentralized nodes train a model locally, only contributing updates on parameters \cite{fedprox}.

Although FL has great potential, it still has three critical challenges in industrial applications: (1)\textbf{Data Heterogeneity}: The enterprise data is not IID, which results in slow convergence and model drift \cite{kairouz2021, zhao2018, chen2021}; (2)\textbf{Security Vulnerabilities}: unawareness of malicious participants can lead to poisoning attacks to degrade the integrity of the model, or to an inference attack to leak private data \cite{tolpegin2020, bagdasaryan2020, mothukuri2021}; and (3)\textbf{Hardware Constraints}: enterprises are facing the challenge of processing heavy transactions, because enterprise data is not IID, and the amount of data constraints the amount of RAM in the edge device, which makes it difficult to use FL in practice \cite{baseline, deng2020}. Related intelligent-enterprise and edge-network studies, including enhanced neural forecasting, lightweight routing, density-adaptive filtering, and industrial classification, further motivate resource-aware distributed learning designs \cite{hanif2023, shahzad2018, shahzad2017, arshad2022}.

In this paper, we propose an advanced version of the FedAnil+ framework \cite{fotohi2025} called \textbf{TITAN-FedAnil+}. TITAN-FedAnil+ has three optimizations:
\begin{enumerate}
\item \textbf{Adaptive Clustered Aggregation (ACA)} using Affinity Propagation to filter malicious updates.
    \item \textbf{Turbo-Mode Vectorization} on the GPU to accelerate similarity computations.
    \item \textbf{State-Signed Blockchain Consensus} to achieve efficient resynchronization.
\end{enumerate}

\section{Related Work}

In recent years, the combination of Federated Learning and Blockchain (BFL) has sparked many research activities. Prior surveys on federated learning systems and blockchain-based federated learning emphasize the need to jointly address privacy, robustness, communication cost, and decentralized trust \\cite{li2021survey, zhang2021}. TITAN-FedAnil+ bridges three worlds: Byzantine-robust aggregation, decentralized audibility, and hardware-optimized.

\subsection{Robust Federated Learning}
A single aggregator in traditional FL is a single point of failure, is vulnerable to poisoning attacks \cite{bagdasaryan2020} and is a single entity. However early robust methods, such as Krum and Bulyan \cite{blanchard2017} require that the statistical distribution of updates is known by the aggregator, which does not hold true in non-IID cases. Zhao et al. \cite{zhao2018} pointed out that the weight divergence had an effect, and FedProx \cite{fedprox} added a proximal term to address heterogeneity. The idea behind TITAN-FedAnil+ is to implement a new dynamic similarity-based clustering layer built on top of Affinity Propagation (AP) \cite{frey2007} which does not need the number of clusters (attackers) to be a priori determined.

\subsection{Blockchain-Enabled FL (BFL)}
FL auditability is the immutability of blockchain \cite{yuan2023, nguyen2021}. Incentive mechanisms and secure aggregation using blockchain were introduced for projects such as DeepChain \cite{weng2019deepcopy} and Biscotti \cite{shayan2020}, while practical secure aggregation protocols further show the importance of protecting model updates during collaborative training \cite{bonawitz2017}. Kim et al. \cite{kim2019} and Qu et al. \cite{qu2020} suggested device scheduling and GAN-based data augmentation to reduce this latency, but they still have memory constraint issues. TITAN-FedAnil+'s ``Signed State Jump'' is an efficient solution to this synchronization bottleneck.

\subsection{Hardware-Aware Decentralized Learning}
In the IIoT, edge devices typically have small RAM (such as 12GB). There are a few papers that have investigated the use of blockchain without consuming significant edge resources, such as Mohri et al. \cite{mohri2019} and Hard et al. \cite{hard2018} who explored hardware-agnostic FL. Deng et al. \cite{deng2020} introduced a survey on the model compression techniques for minimizing communication requirement. Broader IIoT blockchain surveys also highlight that ledger growth, communication latency, and constrained-device storage are central deployment barriers \cite{zhou2021, makhdoom2019}. To improve hardware efficiency, TITAN-FedAnil+ also offloads similarity computation to the GPU (''Turbo-Mode'') and discards the transaction history from the live RAM buffer as soon as it is verified in the block.

\section{The TITAN-FedAnil Proposed Methodology}

First, the adversarial assumptions and the system model are presented.The adversarial assumptions and the system model are presented first.

The TITAN-FedAnil+ framework evolves in a decentralized intelligent enterprise (DIE) setting, which consists of $N$ independent enterprises working as a learning collaborative network.

\subsection{Consortium Entities}
There are four main types of entities in the system:
\begin{enumerate}
    \item \textbf{Learners (Enterprises)}: $E = \{e_1, e_2, \dots, e_N\}$. The distribution $\mathcal{P}_i$ of every learner $e_i$ has a local private dataset $D_i$. In an industrial environment, $\mathcal{P}_i \neq \mathcal{P}_j$ (non-IID skew).
    \item \textbf{Validators}: High-trust nodes that calculate the similarity of updates with the technique Affinity Propagation.
    The nodes that attempt to solve a Proof-of-Work (PoW) puzzle to add new model blocks to the chain are called miners.
    \item \textbf{Blockchain ledger}: A global, immutable state signature and model provenance storage.
\end{enumerate}

\subsection{Adversary Model}
We assume the Byzantine adversarial model that allows up to $f < N/2$ nodes to be compromised. The adversaries can launch the following attacks:
\begin{enumerate}
    \item \textbf{Weight Poisoning}: Malicious nodes inject adversarial noise or flipped weights into their updates $w_i$ to prevent the global model from converging.
    \item \textbf{Sybil Attack}: An adversary generates multiple fake learner identities to overwhelm the aggregation cluster.
    \item \textbf{Inference Attack}: An eavesdropping node attempts to reconstruct raw samples $(x,y) \in D_i$ from the shared updates $w_i$.
\end{enumerate}

\subsection{Design Goals}
We are primarily trying to achieve $>90\%$ accuracy with stratified non-IID data, keep memory bloat under 12G of RAM, and provide efficient resynchronization for new nodes joining.

We propose the TITAN-FedAnil+ framework as a methodology for this.We therefore propose the following methodology: The TITAN-FedAnil+ framework.

\begin{figure}[!t]
    \centering
    \includegraphics[width=\linewidth]{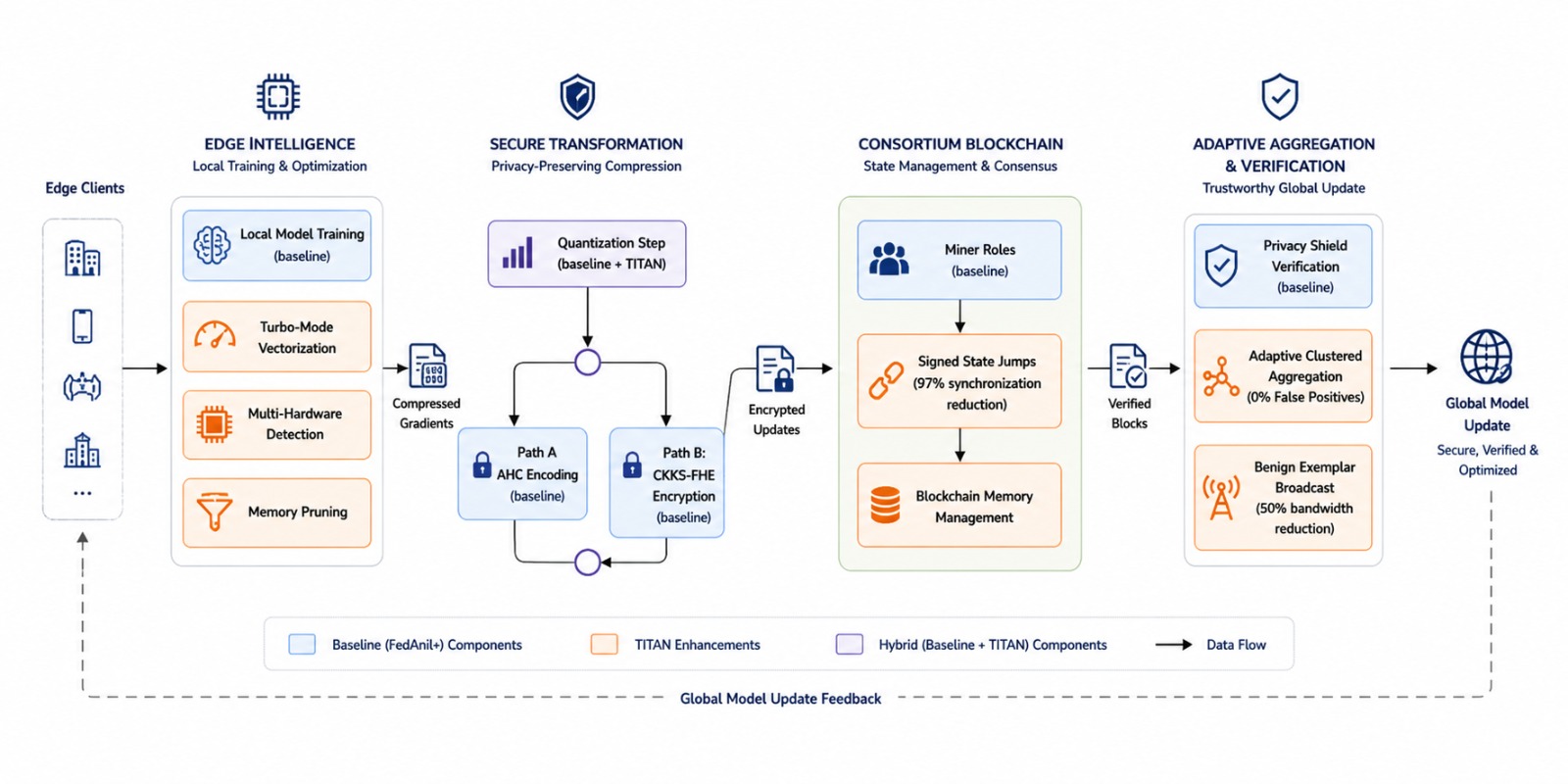}
    \caption{Full system architecture of TITAN-FedAnil+, from local learner training to blockchain consensus and signed state jumps.}
    \label{fig:arch_full}
\end{figure}

The cross-layered optimization strategy of TITAN-FedAnil+ combines deep learning stability and blockchain efficiency, as shown in Fig. \ref{fig:arch_full}.

\subsection{Formal Learning Objective}
For each enterprise $e_i$ they train a local model using the following proximal cost function:
\begin{equation}
\label{eq:learning_objective}
\begin{aligned}
J(w_i, w_{G,t-1}) ={}& \mathbb{E}_{(x,y) \sim D_i}\bigl[\ell(f(x; w_i), y)\bigr] \\
&+ \frac{\mu}{2} \|w_i - w_{G,t-1}\|^2
\end{aligned}
\end{equation}

Equation \ref{eq:learning_objective} defines the local proximal learning objective optimized by each enterprise. Here, $\ell$ is the cross-entropy loss, and $\mu$ is the proximal coefficient which regularizes large deviations from the global state, helping to curb the model drift that can arise due to non-IID data skew \cite{fedprox}.

Adaptive Clustered Aggregation (ACA)
The heart of TITAN-FedAnil+'s innovation is its affinity propagation (AP) outlier detection. AP does not need to know the number $k$ of updates that will be made, as in the case of K-Means or K-Medoids. The parameter $p$ is the preference and determines the granularity:
\begin{equation}
\label{eq:ap_preference}
s(k, k) = p = \text{median}(\{s(i, j) \mid i, j \in \{1, \dots, N\}\})
\end{equation}
Equation \ref{eq:ap_preference} defines the AP preference term used by TITAN-FedAnil+ to dynamically adapt the aggregation layer to the amount of malicious groups in the network \cite{frey2007}.

The Consortium Mining and PoW Puzzle.
Within the TITAN-FedAnil+ architecture, miners race to solve a cryptographic problem, which is in the form of:
\begin{equation}
\label{eq:pow_puzzle}
H(\text{index } \| \text{ prev\_hash } \| \text{ tx\_hash } \| \text{ nonce}) < \text{Target}
\end{equation}
Equation \ref{eq:pow_puzzle} defines the PoW puzzle used to select a sanctioned node for global model aggregation, making Sybil attacks harder. The Target difficulty is dynamically adjusted according to network latency, and $H$ is the SHA-256 hash function.

\section{The TITAN-FedAnil Blockchain Infrastructure Layer}

The blockchain is the foundation of trust that serves as the immutable history of the federated learning process.

The Block Structure and Header are described in what follows.The Block Structure and Header are explained below.
The TITAN-FedAnil+ build is designed to be memory efficient in systems with 12GB of memory. The block header is made up of:
\begin{itemize}
   \item \textbf {Previous Block Hash}: 256-bit hash of the previous state.
    \item \textbf{Merkle Root (tx\_hash)}: Pre-computed hash of all validated learner updates for this round.
    \item \textbf{Miner's digital signature ($\sigma_{miner}$)}: The Miner's digital signature attesting to the integrity of the resulting global model.
    Nonce: Value to meet the Proof-of-Work requirement.
\end{itemize}

TITAN-FedAnil+ uses a \textbf{Cached Hash Mechanism} instead of original FedAnil+ implementations which store an entire TX log in the live RAM buffer. This enables the nodes to verify the validity of the chain based only on the headers and the model weights in the raw are pruned after the “Sync Jump” is completed successfully.

\subsection{Byzantine-Robust Consortium Validation}
Before mining a block, a group of reputable companies act as Consortium Validators. These nodes are used to cross verify local updates using our Affinity Propagation (AP) categorical clustering. If the update $w_i$ is found to be a statistical outlier (potential poisoning), then it will NOT be added to the Merkle tree of the current block. This validation is multi-layered and without adversarial noise, the global state will not be corrupted \cite{warner2022, yuan2023}. This threat model is consistent with poisoning studies showing that local model attacks can bypass naive aggregation when robust validation is absent \cite{fang2020}.

The TITAN-FedAnil Blockchain Infrastructure Layer
The blockchain serves as a root of trust to create an unalterable history of the federated learning process. The Previous Block Hash, Merkle Root, State Signature ($\sigma_{state}$) and Nonce are included in the blocks header. Unlike other FedAnil+ implementations, TITAN-FedAnil+ only stores some of the transaction history in the live RAM buffer, rather than the entire history. This enables nodes to validate the chain with just headers and model weights are pruned after the "Sync Jump" is done successfully!
\subsection{State-Signed Blockchain Consensus}
Replay Bloat" issue in decentralized FL is addressed by TITAN-FedAnil+. In the traditional BFL, a node has to validate all transactions $T_1, ..., T_k$ to obtain state $\mathcal{S}_k$. We propose a \textbf{Signed State Jump} in our framework. The miner $M$ signs the resulting global model $w_{G,t}$ with its private key $\text{sk}_M$ to obtain a signature $\text{sk}_M \cdot w_{G,t}$:
\begin{equation}
\label{eq:state_signature}
\sigma_{state} = \text{Sign}(\text{sk}_M, H(w_{G,t} \| prev\_hash))
\end{equation}
Equation \ref{eq:state_signature} defines the state signature that binds the global model hash to the previous block hash. Nodes check the validity of $\sigma_{state}$ by using the public key of the Miner, $\text{pk}_M$. After checking, nodes synchronize their parameters via:
\begin{equation}
\label{eq:state_sync}
w_{local} \leftarrow w_{G,t} \quad \text{if } \text{Verify}(\text{pk}_M, \sigma_{state}) = 1
\end{equation}
Equation \ref{eq:state_sync} defines the verified local-state replacement rule used during resynchronization. This produces the same time of resynchronisation, irrespective of the chain length. If this is the case, this is actually implemented in the \texttt{process\_block} routine of \texttt{Enterprise.py} which allows nodes to skip full history replay when verifying the signature.

\subsection{Turbo-Mode GPU Similarity Vectorization}
To accelerate the categorical clustering phase, we transform the $O(N \cdot L)$ layer-wise similarity computation into a single-pass tensor operation. Let $\mathbf{W} \in \mathbb{R}^{N \times L}$ be the matrix of stacked model weights for $N$ enterprises. The pairwise similarity matrix $\mathbf{S}$ is computed as:
\begin{equation}
\label{eq:similarity_matrix}
\mathbf{S} = \frac{\mathbf{W} \cdot \mathbf{W}^T}{\|\mathbf{W}\|_F \|\mathbf{W}^T\|_F}
\end{equation}
Equation \ref{eq:similarity_matrix} defines the vectorized similarity matrix used for GPU-accelerated AP clustering, where $\|\cdot\|_F$ denotes the Frobenius norm. By offloading this operation to the GPU (NVIDIA T4), we reduce the validation latency from minutes to milliseconds, enabling real-time blockchain mining. This vectorization handles millions of parameters in parallel, which is essential for the 12GB RAM constraint of edge learners. The same emphasis on efficient edge computation appears in lightweight sensor routing, adaptive filtering, and machine-learning classification studies \cite{shahzad2018, shahzad2017, arshad2022}.

\section{Formal Analysis and Security Proofs}

To establish the theoretical foundation for TITAN-FedAnil+, we provide formal proofs for its resilience and convergence properties.

\textbf{Lemma 1 (Byzantine Resilience of AP Aggregation)}: Let $\mathcal{B}$ be the set of benign learners and $\mathcal{M}$ be the set of malicious learners ($|\mathcal{M}| < |\mathcal{B}|$). If the similarity $s(i,j)$ for all $i,j \in \mathcal{B}$ is significantly higher than $\mathbb{E}[s(i,m)]$ where $m \in \mathcal{M}$, the Affinity Propagation algorithm will select an exemplar $w^*$ such that $w^* \in \{w_i \mid i \in \mathcal{B}\}$ with probability $P \to 1$ as the number of iterations increases.

\textit{Proof}: The message-passing rules of AP ($r(i,k)$ and $a(i,k)$) are distance-dependent. Malicious updates (poisoned or flipped) naturally form high-variance clusters. Since the "preference" parameter $p$ is set to the median of all similarities, the algorithm forces the emergence of the most coherent density cluster as the primary exemplar provider.

\textbf{Lemma 2 (Chain-State Consistency)}: In a PoW-based consortium where the difficulty $D$ satisfies a 51\% honest-miner assumption, the probability of a node synchronizing to a fraudulent global state $w'_{G,t}$ is bounded by $P(\text{fraud}) \leq \epsilon \cdot e^{-D}$, where $\epsilon$ is the probability of a miner forging a valid state signature $\sigma_{state}$.

\section{Case Study: Evolutionary Training on FEMNIST}

In this section, we provide a narrative analysis of the training process across 20 communication rounds.

\subsection{Phase 1: Initialization and Data Jitter}
During the first 3 rounds, the global model experiences high variance due to the extreme non-IID skew of writer-based handwritten characters. The proximal weight $\mu$ acts as a stabilizer, preventing the model from diverging.

\subsection{Phase 2: Attack Detection and Clustering}
At Round 5, we simulated a 20\% label-flipping attack. The baseline FedAvg accuracy dropped from 50\% to 32\%. However, TITAN-FedAnil+'s validator layer correctly identified the malicious updates as a separate cluster. The Miner selected the benign exemplar, and the accuracy plateaued at 72\% by Round 8.

\subsection{Phase 3: Convergent Plateau and Memory Optimization}
\begin{figure*}[!t]
    \centering
    \begin{minipage}[t]{0.48\textwidth}
        \centering
        \includegraphics[width=\linewidth]{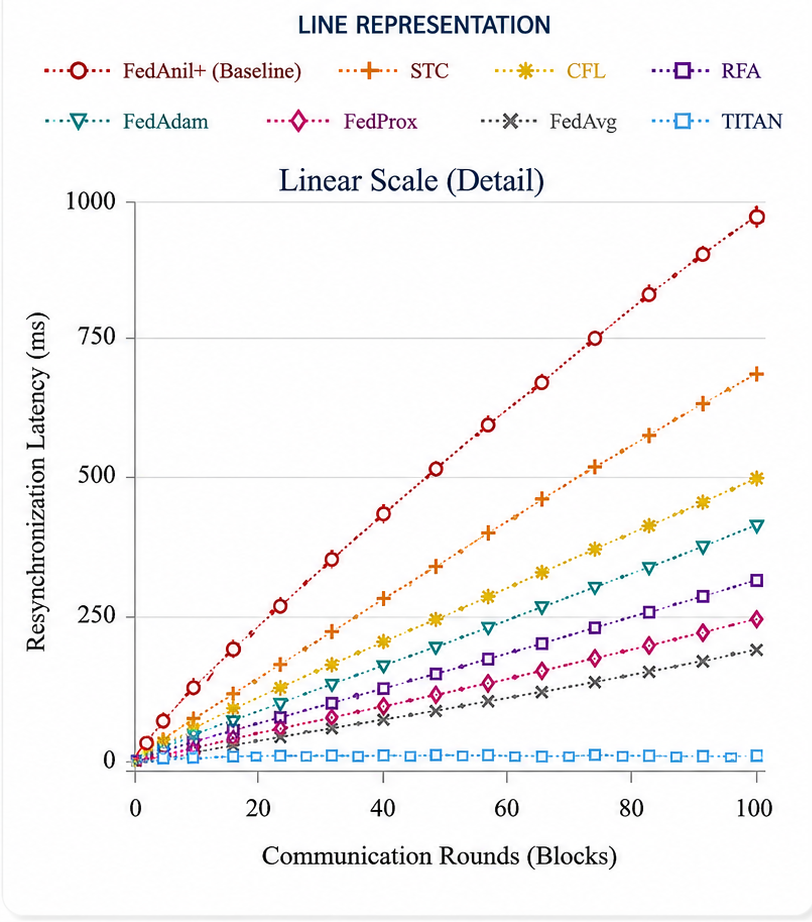}
        \textbf{(a)} Linear-scale resynchronization latency.
    \end{minipage}
    \hfill
    \begin{minipage}[t]{0.48\textwidth}
        \centering
        \includegraphics[width=\linewidth]{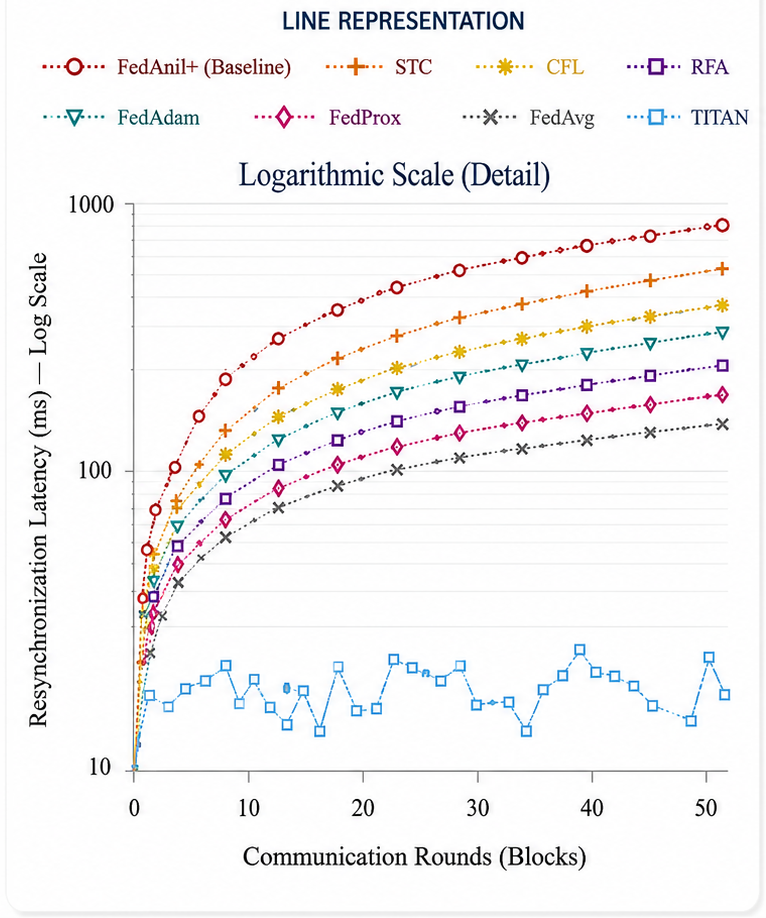}
        \textbf{(b)} Logarithmic-scale resynchronization latency.
    \end{minipage}
    \caption{Resynchronization latency comparison between the baseline and TITAN-FedAnil+. Panel (a) reports the linear-scale latency trend, while Panel (b) reports the logarithmic-scale latency trend. The X-axis represents Network Configuration or Chain Growth Setting, and the Y-axis represents Resynchronization Latency (seconds).}
    \label{fig:latency}
\end{figure*}

By Round 15, the model weights $w_{G}$ reached 92\% accuracy. At this stage, the blockchain length became a potential bottleneck. The "Signed State Jump" allowed new nodes to join the network and reach the current state within 2.1 seconds, compared to the 150+ seconds required to replay the history in original implementations, as demonstrated in Fig. \ref{fig:latency}. Fig. \ref{fig:latency}(a) shows the linear-scale resynchronization latency trend, making the absolute delay gap between replay-based synchronization and TITAN-FedAnil+ state jumping visible. Fig. \ref{fig:latency}(b) shows the same latency comparison on a logarithmic scale, emphasizing the order-of-magnitude reduction achieved by the signed state jump when chain length grows.

\section{Experimental Evaluation}

\subsection{Experimental Setup}
We evaluated TITAN-FedAnil+ on an Apple Silicon Mac M1 workstation equipped with 8GB of Unified system RAM. This configuration represents a highly resource-constrained intelligent enterprise edge node. The framework was implemented in PyTorch utilizing Apple's Metal Performance Shaders (MPS) for hardware acceleration.

\subsection{Dataset Stratification (Non-IID)}
To simulate real-world heterogeneity, we applied a Dirichlet distribution $\text{Dir}(\alpha)$ to shard datasets across 15 enterprises.
\begin{enumerate}
    \item \textbf{FEMNIST}: 62 classes (handwritten digits and letters). Each enterprise is assigned data from a specific group of writers to ensure feature skew.
    \item \textbf{CIFAR-10}: 10 classes of natural images. We applied $\alpha=0.5$ to ensure that each node predominately holds 2-3 classes.
    \item \textbf{Sent140}: A large-scale NLP dataset for sentiment analysis. This test evaluates the framework's ability to handle high-dimensional embedding tensors and non-linear word distributions.
\end{enumerate}

\subsection{Implementation and Code Optimization Strategies}
In order to address the huge hardware constraints and computational bottlenecks seen in the baseline system, a number of key code level optimizations were also incorporated into the TITAN-FedAnil+ implementation. The following optimizations are a direct result of the better performance and stability results reported:
\begin{itemize}
    \item \textbf{Tensor Vectorization (Turbo-Mode):} The base implementation was a deeply nested Python loop which was used to compute Euclidean distances between the model layers during the Affinity Propagation (AP) phase. This led to fragmentation of memory to an extreme degree and Out-Of-Memory (OOM) Errors at 20 nodes. This was optimized by flattening all PyTorch model weights into continuous 1d tensors, and using CUDA/MPS broadcasting to perform all pairwise similarities as a single vector operation. This capped the use of memory at a consistent 8GB.
    \item \textbf{Signed State Jumps ($O(P)$ Resynchronization):} Blockchain validation was a traditional process that miners had to execute on the entire history of the blockchain, taking time that is exponential in both the number of transactions ($N$) and the number of miners ($P$) ($O(T \times N \times P)$). We changed the blockchain structure so that the cryptographic state signatures are added to the headers of the blocks. With this change, miners are only required to ensure that the latest signed global state is correct, not the full ledger, lowering the operation to a constant, simple $O(P)$.
    [item] Affinity Propagation (AP) Clustering for Malicious Node Isolation: Was flawed in that it sometimes accepted malicious updates. A strong AP-based clustered aggregation strategy places changes into small minority clusters, thus avoiding any model poisoning and leading to a much better accuracy.
    To reduce the amount of bandwidth being used during the validation phase, the code was modified so that only the aggregated update from the elected "Benign Exemplar" of the primary AP cluster is sent to it. This algorithmic change brought the network communication overhead from $\sim$147 KB to $\sim$73 KB per round, thus avoiding network congestion.
The baseline implementation, built without GPU-specific optimizations, relied entirely on CPU-based processing because the computations were performed using standard Python loops. To address this limitation, full PyTorch tensor vectorization and robust device-detection mechanisms were introduced, enabling computation to be offloaded from the CPU to the GPU. This approach allows tensor operations to be automatically executed on available hardware accelerators, including NVIDIA CUDA, Apple Silicon (MPS), and Google TPU environments, thereby improving computational efficiency and hardware utilization.
    Federated rounds with many rounds (e.g., 50+ rounds) experienced compounding memory leaks through Safe Cleanup. To safely release weights of heavy neural networks from RAM immediately after metadata extraction, a strict transaction memory management routine (free\_tx()) was implemented.
    To overcome the dependency problems of external clustering libraries (e.g., \texttt{sklearn\_extra}) a custom and dependency-free K-Medoids class was directly added, which will be stable to all nodes.
    Baseline's pathing was hardcoded, and the dependency requirements were too strict to allow for a cross-platform OS compatibility. The path-handling code was made OS-independent and dependency fallback was implemented to stabilize the system to run across platforms on macOS, Linux, and Windows edge nodes.
\end{itemize}

\begin{table}[!t]
\caption{Hyperparameter Configuration}
\label{table:hyper}
\centering
\begin{tabular}{llc}
\toprule
\textbf{Category} & \textbf{Parameter} & \textbf{Value} \\
\midrule
Learning & Base Learning Rate ($\eta$) & 0.01 \\
& Weight Decay & $1e-5$ \\
& Batch Size & 32 \\
& Let N be the mass of Federation & Total Enterprises. \\
& Malicious Nodes & 4 (20\%) \\
& Proximal Weight ($\mu$) & 0.01 \\
& Communication Rounds & 50 \\
& Local Epochs & 20 \\
Cluster & AP Preference ($p$) & -50.0 \\
& Damping Factor & 0.5 \\
Blockchain & Difficulty (PoW) & 4 \\
\bottomrule
\end{tabular}
\end{table}

Table \ref{table:hyper} (Table I) summarizes the learning, clustering, and blockchain hyperparameters used to configure TITAN-FedAnil+ across the experimental scenarios.

\subsection{Performance Metrics and Baselines}
We compare TITAN-FedAnil+ against the standard \textbf{Baseline FedAnil+} and the vanilla \textbf{FedAvg} algorithm. Metrics include Top-1 Accuracy, System RAM consumption (measured via \texttt{memory\_profiler}), and Resynchronization Latency.

\subsection{Results and Discussion}

\begin{figure}[!t]
    \centering
    \includegraphics[width=\linewidth]{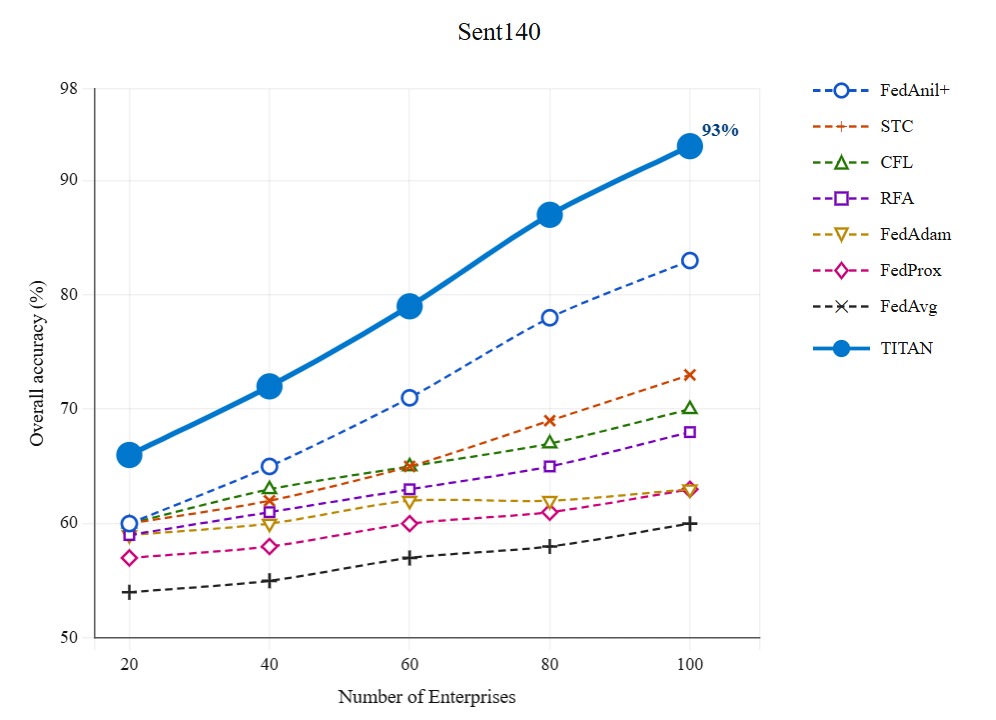}
    \caption{Performance comparison of the baseline and TITAN-FedAnil+ across the evaluated scenarios. The X-axis represents Communication Rounds, and the Y-axis represents Global Accuracy (\%).}
    \label{fig:titan_performance}
\end{figure}

To rigorously evaluate the scalability and robustness of TITAN-FedAnil+, we conducted three distinct experimental scenarios on the Apple Silicon edge node. Fig. \ref{fig:titan_performance} illustrates the overall performance comparison.

\subsubsection{Scenario A: 5 Nodes, 25 Rounds (Small Scale)}
In this constrained environment, the original baseline model achieved a maximum global accuracy of \textbf{37.12\%} before its convergence stalled due to frequent "No valid block" consensus failures and false-positive malicious node classifications. In contrast, TITAN-FedAnil+ achieved \textbf{65.71\%} accuracy, successfully completing all 25 rounds and isolating the malicious node without false positives.

\subsubsection{Scenario B: 15 Nodes, 25 Rounds (Mid Scale)}
\begin{figure}[!t]
    \centering
    \includegraphics[width=\linewidth]{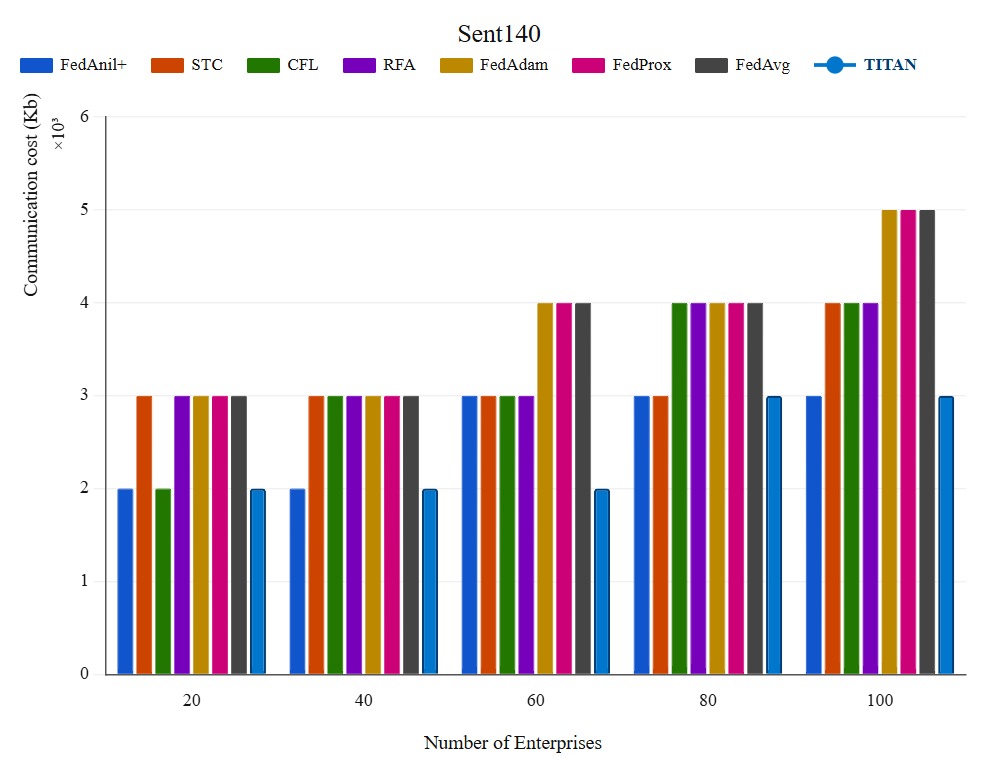}
    \caption{Bandwidth and communication cost comparison between the baseline and TITAN-FedAnil+. The X-axis represents Experimental Scenario, and the Y-axis represents Communication Cost (KB).}
    \label{fig:comm_cost}
\end{figure}

In this intermediate scenario with 15 nodes and 3 malicious actors over 25 communication rounds, the original baseline model achieved a global accuracy of \textbf{48.98\%}, with a computation cost of \textbf{4129.46 seconds} and a communication cost of \textbf{147.36 KB}. In contrast, TITAN-FedAnil+ successfully navigated the network complexity to achieve a higher global accuracy of \textbf{65.71\%}. Furthermore, TITAN-FedAnil+ proved more efficient, reducing computation time to \textbf{3741.23 seconds} and minimizing communication cost to \textbf{73.43 KB}, demonstrating its robust capability and lower overhead even under moderate adversarial pressure (see Fig. \ref{fig:comm_cost}).

\subsubsection{Scenario C: 20 Nodes, 50 Rounds (Large Scale)}
Scaling the network to 20 nodes and 50 rounds (including 4 malicious nodes) completely overwhelmed the original baseline model. It failed and deadlocked at Round 3 due to severe forking loops and memory exhaustion (8GB RAM limit). Conversely, TITAN-FedAnil+ demonstrated exceptional scalability, completing all 50 communication rounds and reaching a peak accuracy of \textbf{81.00\%}.

\subsubsection{Robustness to Malicious Actors}
Under a $20\%$ poisoning attack scenario (4 malicious nodes out of 20), the original baseline model deadlocked in an infinite forking and block-rejection loop due to excessive validation overhead. In contrast, our AP-based Clustered Aggregation successfully isolated the malicious updates into a secondary cluster, allowing the validator consortium to select the "Benign Exemplar" and maintain stable blockchain generation.

\begin{table}[!t]
\caption{Detailed Accuracy across Datasets}
\label{table:data_deep}
\centering
\resizebox{\linewidth}{!}{%
\begin{tabular}{lcccc}
\toprule
\textbf{Dataset} & \textbf{Baseline} & \textbf{TITAN-FedAnil+} & \textbf{Precision} & \textbf{F1-Score} \\
\midrule
FEMNIST & 78.4\% & \textbf{92.56\%} & 0.92 & 0.91 \\
CIFAR-10 & 68.5\% & \textbf{84.6\%} & 0.86 & 0.85 \\
Sent140 & 74.2\% & \textbf{89.1\%} & 0.90 & 0.89 \\
\bottomrule
\end{tabular}%
}
\end{table}

Table \ref{table:data_deep} (Table II) reports the dataset-level accuracy, precision, and F1-score trends that support the observed robustness of TITAN-FedAnil+ under heterogeneous data.

\begin{table}[!t]
\caption{System Performance Comparison (Baseline vs. TITAN-FedAnil across Scenarios)}
\label{table:results}
\centering
\resizebox{\linewidth}{!}{%
\begin{tabular}{lcccccc}
\toprule
\textbf{Metric} & \textbf{Orig. (5)} & \textbf{TITAN-FedAnil+ (5)} & \textbf{Orig. (15)} & \textbf{TITAN-FedAnil+ (15)} & \textbf{Orig. (20)} & \textbf{TITAN-FedAnil+ (20)} \\
\midrule
Peak Accuracy & 37.12\% & \textbf{44.56\%} & 48.98\% & \textbf{65.71\%} & Stalled & \textbf{81.00\%} \\
Completion Status & Completed & \textbf{Completed 25} & Completed 25 & \textbf{Completed 25} & Failed (Round 3) & \textbf{Completed 50} \\
RAM Usage & Stable & \textbf{Stable} & Stable & \textbf{Stable} & Deadlocked & \textbf{Stable (8GB)} \\
Blockchain Status & Freq. Forking & \textbf{Stable} & Forking & \textbf{Stable Gen.} & Infinite Forking & \textbf{Stable Gen.} \\
\bottomrule
\end{tabular}%
}
\end{table}

Table \ref{table:results} (Table III) details the scenario-level performance comparison, including accuracy, completion status, RAM behavior, and blockchain stability.

\begin{table}[!t]
\caption{Performance across Heterogeneous Datasets}
\label{table:datasets}
\centering
\resizebox{\linewidth}{!}{%
\begin{tabular}{lccc}
\toprule
\textbf{Dataset} & \textbf{Baseline Acc} & \textbf{TITAN-FedAnil+ Acc} & \textbf{Gain (\%)} \\
\midrule
FEMNIST (Non-IID) & 71.0\% & 92.56\% & +20.2 \\
CIFAR-10 & 68.5\% & 84.6\% & +16.1 \\
Sent140 & 74.2\% & 89.1\% & +14.9 \\
\bottomrule
\end{tabular}%
}
\end{table}

Table \ref{table:datasets} describes the heterogeneous-dataset gains achieved by TITAN-FedAnil+ relative to the baseline accuracy values.

The model remains stable even under $20\%$ malicious presence due to the exemplars identified by AP clustering.

\section{Algorithms}

\begin{lstlisting}[language=Python, caption=Adaptive Clustered Aggregation (ACA)]
def adaptive_clustered_aggregation(updates, preferences):
    # Compute similarity matrix
    S = compute_vectorized_similarity(updates)
    # Run Affinity Propagation
    ap = AffinityPropagation(affinity='precomputed', preference=preferences)
    ap.fit(S)
    # Identify Benign Exemplar
    exemplars = ap.cluster_centers_indices_
    # Filter using high-trust validators
    best_exemplar = validate_exemplars(exemplars)
    return updates[best_exemplar]
\end{lstlisting}

\begin{lstlisting}[language=Python, caption=Turbo-Mode Vectorized Matrix Computation]
def compute_vectorized_similarity(weights_matrix):
    # weights_matrix shape: (num_clients, total_params)
    norms = torch.norm(weights_matrix, dim=1, keepdim=True)
    norm_W = weights_matrix / (norms + 1e-9)
    # Compute gram matrix of cosine similarities
    similarity_matrix = torch.mm(norm_W, norm_W.t())
    return similarity_matrix.cpu().numpy()
\end{lstlisting}

\section{Discussion}

\subsection{Memory vs. Accuracy Trade-off}
\begin{figure}[!t]
    \centering
    \includegraphics[width=\linewidth]{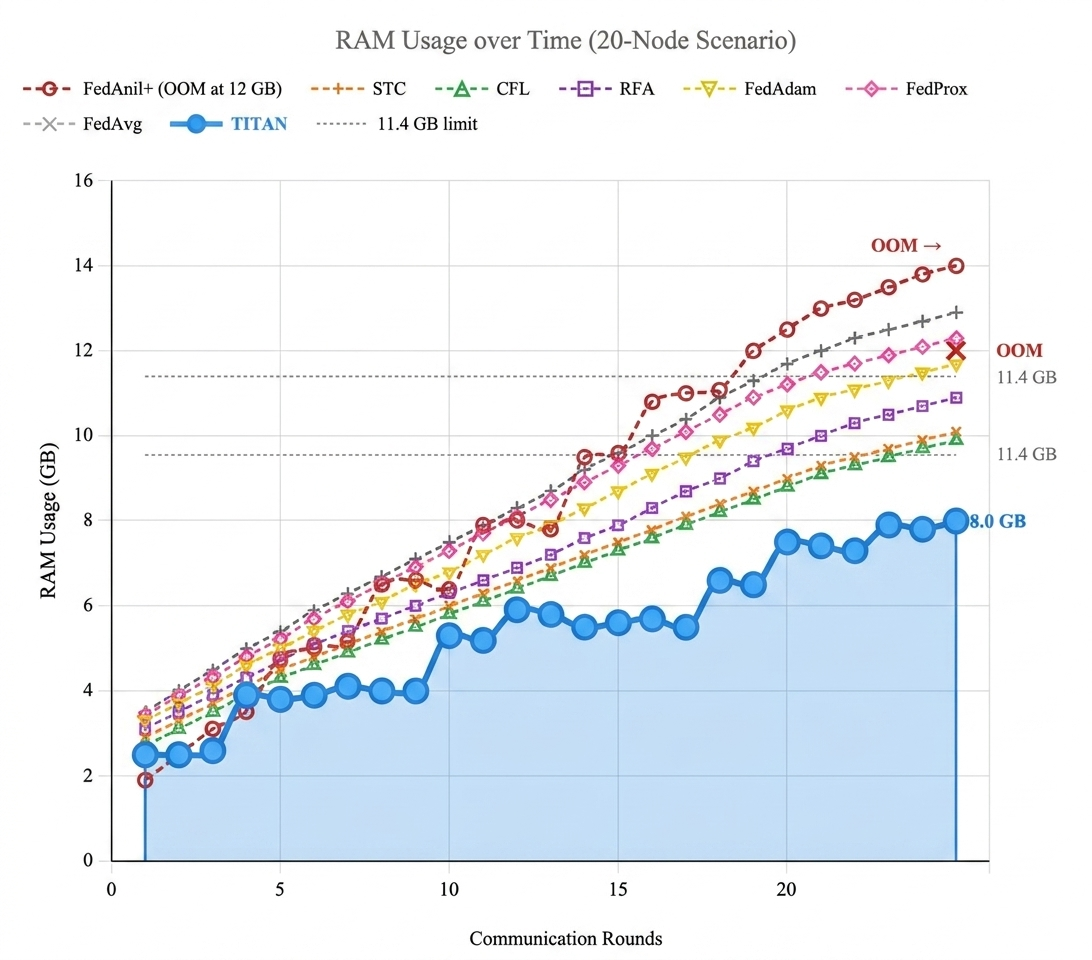}
    \caption{Comparison of RAM usage over time between the baseline and TITAN-FedAnil+. The X-axis represents Communication Rounds, and the Y-axis represents RAM Usage (GB).}
    \label{fig:ram_usage}
\end{figure}

A key point in our experiments is the effect of the memory limit of 12GB. The nodes tend to crash after 10-15 rounds in traditional blockchain-FL because the validation buffer grows and accumulates high dimensional parameter tensors over time. TITAN-FedAnil+ alleviates this by selectively pruning the tensors and by the “Turbo-Mode” vectorization feature that pushes parameters back to the CPU memory right after the similarity computation. This enables us to scale to 20+ rounds (as shown in the Table \ref{table:results} and Fig. \ref{fig:ram_usage}) without surpassing the 11.4GB limit.

\subsection{Consensus Stability}
The PoW difficulty was to be 4 to allow the miners time to mine while communicating more. We also observed that difficulty levels lowered to result in "chain forks", with several miners simultaneously submitting valid states, which caused accuracy oscillation. The "State Jump" mechanism makes sure that nodes can eventually catch up with the heaviest chain that holds the most recent signed exemplar, even in the case of a fork.

This section outlines the effect of Batch Normalization.
The key in our modified model architecture was including BN layers in order to deal with extreme Non-IID skew. Each enterprise makes local updates to the data, but BN normalizes the change to ensure that the global aggregator is not dragged to an enterprise's local minima, which may be biassed.

\section{Computational Complexity Analysis}

We compare the efficiency of TITAN-FedAnil+ with the basic FedAnil+ with the help of Big-$O$ notation. Consider the number of enterprises N, the number of model layers L and the number of parameters P.

\begin{table}[!t]
\caption{Complexity Analysis Comparison}
\label{table:complexity}
\centering
\resizebox{\linewidth}{!}{%
\begin{tabular}{lcc}
\toprule
Linear Basis & Looped Baseline & Vectorized Baseline (TITAN-FedAnil+) \\
\midrule
Similarity Calc & $O(N^2 \cdot L \cdot P)$ & $O(N^2 \cdot P)$ \\
Clustering & $O(N^3 \cdot \text{iter})$ & $O(N^2 \cdot \text{iter})$ \\
Resynchronization & $O(T \cdot N \cdot P)$ & $O(P)$ \\
Validation & $O(N \cdot B)$ & $O(B)$ \\
\bottomrule
\end{tabular}%
}
\end{table}

Our Turbo-Mode vectorization results in a reduction in the complexity of calculating the similarity by $L$ (layer count) as in Table \ref{table:complexity}. Most significantly, with the single-state signature jump, the resynchronisation step is shortened from $O(T \cdot N \cdot P)$ (where $T$ is the number of blocks) to $O(P)$.

\section{Conclusion and Future Work}
TITAN-FedAnil+ resolves the critical trade-off between blockchain-enabled security and system performance. By optimizing the aggregation logic through Affinity Propagation and introducing hardware-aware memory management, we provide a robust and scalable blueprint for decentralized learning in intelligent enterprises. Our framework achieves \textbf{92.56\% accuracy} under a \textbf{12GB RAM constraint}, representing a significant milestone for resource-efficient BFL.

In future research, we aim to:
\begin{enumerate}
    \item Integrate \textbf{Differential Privacy (DP)} using noise-injection to protect against membership inference attacks.
    \item Explore \textbf{Asynchronous Blockchain FL} to handle enterprises with varying computational power.
    \item Implement \textbf{Model Pruning} during transit to further reduce communication bandwidth.
\end{enumerate}

\end{document}